# An Invisible Metallic Mesh

Dexin Ye[1,2], Ling Lu[2], John D. Joannopoulos[2], Marin Soljačić[2] and Lixin Ran[1]

[1]Laboratory of Applied Research on Electromagnetics (ARE), Zhejiang University, Hangzhou 310027, China

[2]Department of Physics, Massachusetts Institute of Technology, Cambridge, MA 02139, USA

We introduce a solid material that is itself invisible, possessing identical electromagnetic properties as air (i.e. not a cloak) at a desired frequency. Such a material could provide improved mechanical stability, electrical conduction and heat dissipation to a system, without disturbing incident electromagnetic radiation. One immediate application would be towards perfect antenna radomes. Unlike cloaks[1,2], such a transparent and self-invisible material has yet to be demonstrated. Previous research[3-18] has shown that a single sphere or cylinder coated with plasmonic or dielectric layers can have a dark-state with considerably suppressed scattering cross-section, due to the destructive interference between two resonances in one of its scattering channels. Nevertheless, a massive collection of these objects will have an accumulated and detectable disturbance to the original field distribution. Here we overcome this bottleneck by lining up the dark-state frequencies in different channels. Specifically, we derive analytically, verify numerically and demonstrate experimentally that deliberately designed corrugated metallic wires can have record-low scattering amplitudes, achieved by aligning the nodal frequencies of the first two scattering channels. This enables an arbitrary assembly of these wires to be omnidirectionally invisible and the effective constitutive parameters nearly identical to air. Measured transmission spectra at microwave frequencies reveal indistinguishable results for all the arrangements of the 3D-printed samples studied.

This paper is organized as follows. First, we present the analytical solution of an ideal infinite corrugated thin conducting wire having an extremely low scattering width at a particular frequency. Materials made of such wires, having identical permittivity and permeability as air, would be invisible. Second, we design such an invisible medium using realistic materials modeled by full wave simulations.

Lastly, we describe the fabrication of a sample using this design and demonstrate its omnidirectional invisibility using microwave measurements.

We begin by considering the plane wave scattering by an infinite cylindrical conducting wire with a radius $r_1$, as shown in Fig. 1a (when $r_2=r_1$). When the incident electric field is parallel to the wire, the normalized scattering width $\sigma_{sca}/2r_1$ can be derived as $\sigma_{sca}/2r_1 = (2/k_0 r_1)\sum_{n=-\infty}^{+\infty}|J_n(k_0 r_1)/H_n^{(2)}(k_0 r_1)|^2$ by the Mie solution[19] (see section 1 in **supplementary materials**), where $J_n$ and $H_n^{(2)}$ denote a Bessel function and a Hankel function of the second kind, respectively. The normalized scattering width is plotted as the dashed gray line in Fig. 1c. At high frequencies, the scattering width $\sigma_{sca}$ approaches a value equal to twice the wire diameter. There is only one resonance occurring at zero (DC) frequency, where the scattering width diverges. This can be explained using the resonance frequency ($\omega_0=1/(2\pi\sqrt{LC})$) of inductance $L$ and capacitance $C$ of circuit elements. A thin long wire has an effective infinite inductance and capacitance ($L, C \to \infty$), leading to a zero resonance frequency ($\omega_0 \to 0$).

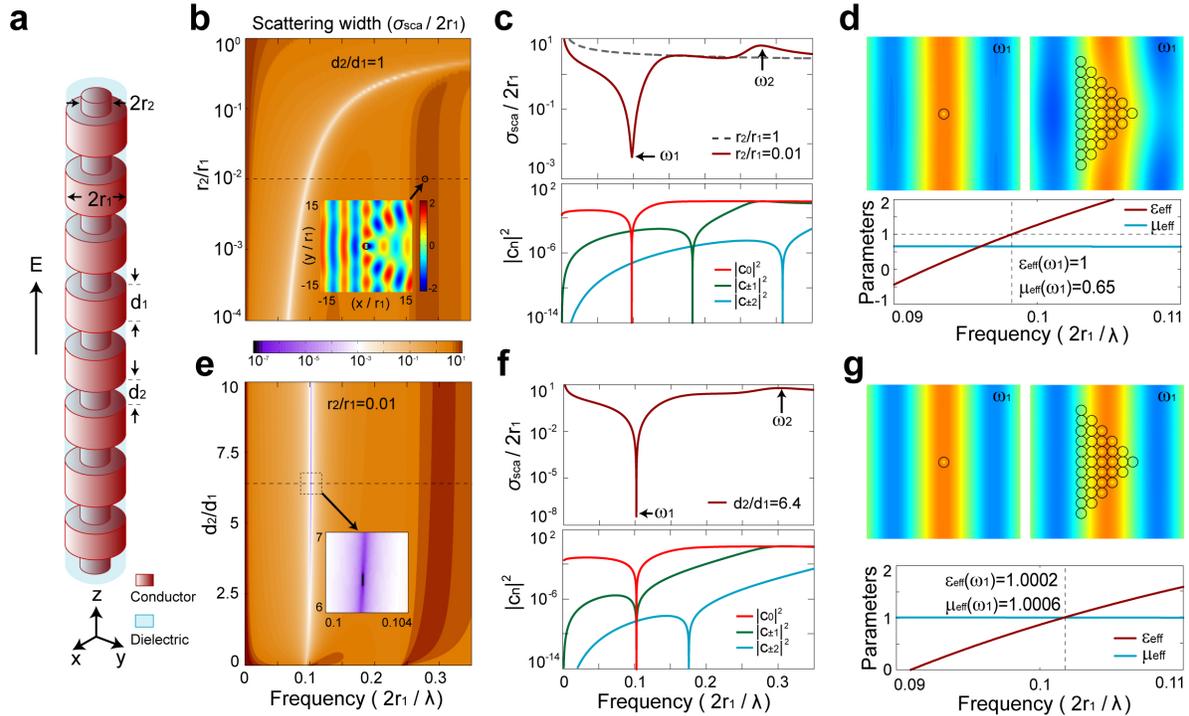

**Figure 1 | Analytical solutions to wave-scattering off corrugated conducting wires. a**, Geometry of a corrugated cylindrical conducting wire, whose open volumes are filled with a dielectric $\varepsilon_r = 6\varepsilon_0$. **b**, The normalized total scattering widths for varying $r_2/r_1$ with subwavelength features $d_1$, $d_2$, set to $d_2/d_1=1$. The inset shows the total electric field around the wire under a unit-amplitude

plane wave incidence at normalized frequency $\omega_2 \approx 0.28$ for $r_2/r_1$ =0.01. **c**, Top: the total scattering widths for $r_2/r_1$ = 0.01(red line) and $r_2/r_1$ =1 (dashed gray line), the minimum scattering width is $4.8\times10^{-3}$ at $\omega_1$. Bottom: the corresponding magnitudes of scattering coefficients ($|c_n|^2$) of different orders. The dips for 0[th], 1[st] and 2[th] orders appear at different frequencies, respectively. **d**, Top left: Total electric field scattering off a single wire. Top right: Total electric field scattering off multiple closely arranged wires. The incident fields are unit-amplitude plane waves at frequency $\omega_1$. Bottom: The retrieved effective constitutive parameters of the closely arranged wires, $\varepsilon_{eff}(\omega_1)$=1 and $\mu_{eff}(\omega_1)$=0.65. **e**, Normalized total scattering width as a function of $d_2/d_1$ with $r_2/r_1$ =0.01. **f**, Top: otal scattering width with $r_2/r_1$ =0.01 and $d_2/d_1$ =6.4, and the minimum normalized scattering width is $3.5\times10^{-8}$ at $\omega_1$ =0.1017. Bottom: Scattering coefficients, where the nodal frequencies of the 0[th] and 1[st] orders coincide at $\omega_1$. **g**, Top left: Total electric field for a single wire. Top right: Total electric fields for multiple closely arranged wires. Bottom: Retrieved effective constitutive parameters of the closely arranged wires, $\varepsilon_{eff}(\omega_1)$=1.0002 and $\mu_{eff}(\omega_1)$=1.0006.

It is known that a scattering dark-state (scattering dip in frequency) universally exists between two resonances (scattering peaks), where the two resonances destructively interfere with equal amplitude but opposite phase in the same scattering and polarization channel[18]. We can thus create such a scattering-dark state by introducing a second resonance in the wire other than the one at DC. To introduce more resonances, we corrugate the wire by shrinking it periodically along the wire as shown in Fig. 1a. This corrugated wire consists of coaxial cylinders with different radii ($r_1$ and $r_2$) and heights ($d_1$ and $d_2$), where $d_1$ and $d_2$ are both far smaller than the free space wavelength. The open volume can be filled with a low-loss material (dielectric constant $\varepsilon_r$) to improve the mechanical strength and increase the working wavelength, so the corrugations are more subwavelength and can be well described by the effective medium theory[20]. Each open cylindrical space inside the wire forms a whispering gallery (WG) resonator; its spectrum and mode profiles are plotted in Fig. S1 in section 3 of **supplementary materials**.

For such an infinite corrugated wire, its normalized scattering width can also be analytically derived, under the effective medium approximation, as (See section 1 of **supplementary materials** for detailed derivation)

$$\sigma_{sca}/2r_1 = \frac{2}{k_0 r_1} \sum_{n=-\infty}^{+\infty} |c_n|^2, \qquad (1)$$

where $c_n = \dfrac{J_n(k_0 r_1)[J_n(k_{eff} r_2)Y_n'(k_{eff} r_1) - J_n'(k_{eff} r_1)Y_n(k_{eff} r_2)] - \eta_{eff} J_n'(k_0 r_1)[J_n(k_{eff} r_2)Y_n(k_{eff} r_1) - J_n(k_{eff} r_1)Y_n(k_{eff} r_2)]}{\eta_{eff} H_n^{2\prime}(k_0 r_1)[J_n(k_{eff} r_2)Y_n(k_{eff} r_1) - J_n(k_{eff} r_1)Y_n(k_{eff} r_2)] - H_n^2(k_0 r_1)[J_n(k_{eff} r_2)Y_n'(k_{eff} r_1) - J_n'(k_{eff} r_1)Y_n(k_{eff} r_2)]}$

denotes the expanded $n$[th]-order scattering coefficient in Bessel ($J_n$ and $Y_n$), Hankel functions ($H_n^{(2)}$), and their derivatives. Here $k_{eff} = \omega n_{eff} = \omega\sqrt{u_{\parallel}^{eff} \times \varepsilon_z^{eff}} = \omega\sqrt{\varepsilon_r u_0}$, $\eta_{eff} = \sqrt{u_{\parallel}^{eff} / \varepsilon_z^{eff}} = [d_2/(d_1+d_2)]\sqrt{u_0/\varepsilon_r}$, where $\varepsilon_z^{eff} = \varepsilon_r(d_1+d_2)/d_2$ and $\mu_{\parallel}^{eff} = \mu_0 d_2/(d_1+d_2)$ are the effective permittivity and permeability of the corrugated volume ($r_2 \leqslant \rho \leqslant r_1$) derived in the **supplementary materials**. We note that $J_0'(k_0 r_1)$ approaches zero when $k_0 r_1$ is a small number for a thin wire. So the nodal frequency of $c_0$ has almost no dependence on $\eta_{eff}$, consequently independent of $d_1$ and $d_2$. The sum of the infinite orders of scattering channels can be terminated to the first few terms, since higher order contributions diminish exponentially[10]. Here we neglect the terms for $|n| > 5$.

Shown as a white stripe in Fig. 1b, there will always exist a scattering dark state whose frequency $\omega_1$ lies between those of the DC resonance and the first WG resonance ($\omega_2 \approx 0.28$) of the corrugated wire. In this plot, we vary the ratio of $r_2/r_1$ while fixing $d_1/d_2 = 1$ and $\varepsilon_r = 6$. In Fig. 1c, we decompose the total scattering width into individual orders and find that each order has its own zero-scattering frequency. Since the 0th order is dominant, the dip in the total scattering corresponds to the nodal frequency of the 0th scattering order $c_0$. The same mechanism enabled previous studies on transparent (cloaking) wires[3-9], invisible particles [10-18] or scattering dark states[21-24].

However, the vanishing of $c_0=0$ is not enough to make a collection of these wires invisible; the total scattering amplitude is not small enough when $c_1 \neq 0$. In Fig. 1d, we show the obvious distortion of a scattered wave by a set of these wires packed closely. For a better understanding, we also show the effective constitutive parameters of an array of such wires at the bottom of Fig. 1d, using a homogenization approach[25]. At $\omega_1$, $\varepsilon_{eff} = 1$ and $\mu_{eff} = 0.65$. Although the effective permittivity of the material $\varepsilon_{eff}$ is 1 (the same as that of free space), $\mu_{eff} \neq 1$. By solving the Mie scattering solutions for a homogeneous dielectric thin wire (See section 2 of supplementary materials), we show that $c_0=0$ requires $\varepsilon=1$ and $c_1=0$ requires $\mu=1$. Since the nodal frequencies of $c_0$ and $c_1$ in general occur at different frequencies for a single element, $\varepsilon=\mu=1$ cannot be satisfied simultaneously for an assembly of them. This is why no transparently invisible metamaterial has been reported to date.

Now, we tune the nodal frequency of $c_1$ to coincide with that of $c_0$ for an individual wire (i.e. $c_0 = c_1 = 0$ at the same frequency), which results in a further decrease of the total scattering amplitude by five orders of magnitude to a negligible value. Consequently, $\varepsilon_{eff} \approx \mu_{eff} \approx 1$ for an arbitrary assembly of such wires. We achieve this by tuning the geometry of the corrugation. We have seen that $c_0$ is almost independent of $d_2/d_1$, while the $c_i$ (i>0) have a strong dependence on $d_2/d_1$. For example, the white line (nodal frequency of $c_0$) in Fig. 1e is almost a straight vertical line that does not change with $d_2/d_1$. So by varying $d_2/d_1$, we can freely tune the nodal frequency of $c_1$ towards that of $c_0$. Starting with the configuration in Fig. 1b where $r_2/r_1$ is fixed at 0.01, we tune the ratio of $d_2/d_1$ from 1 to ~6.4 in Fig. 1e. The nodal frequencies of $c_0$ and $c_1$ coincide and the total scattering width decreases by five orders of magnitude to a record-low scattering width of $3.5\times10^{-8}$ (which eventually will be limited by material losses in experiments). At the same time, $\mu_{eff}$ increases from 0.65 to 1.0006. Consequently, the wave experiences no distortion after impinging on closely arranged wires in Fig. 1g (compared to Fig. 1d). This means arbitrary composites of such wires should be practically invisible. We emphasize that such an alignment of nodal frequencies can robustly occur at any frequency by tuning $r_2/r_1$ and $d_2/d_1$ (see **supplementary materials** Fig. S2).

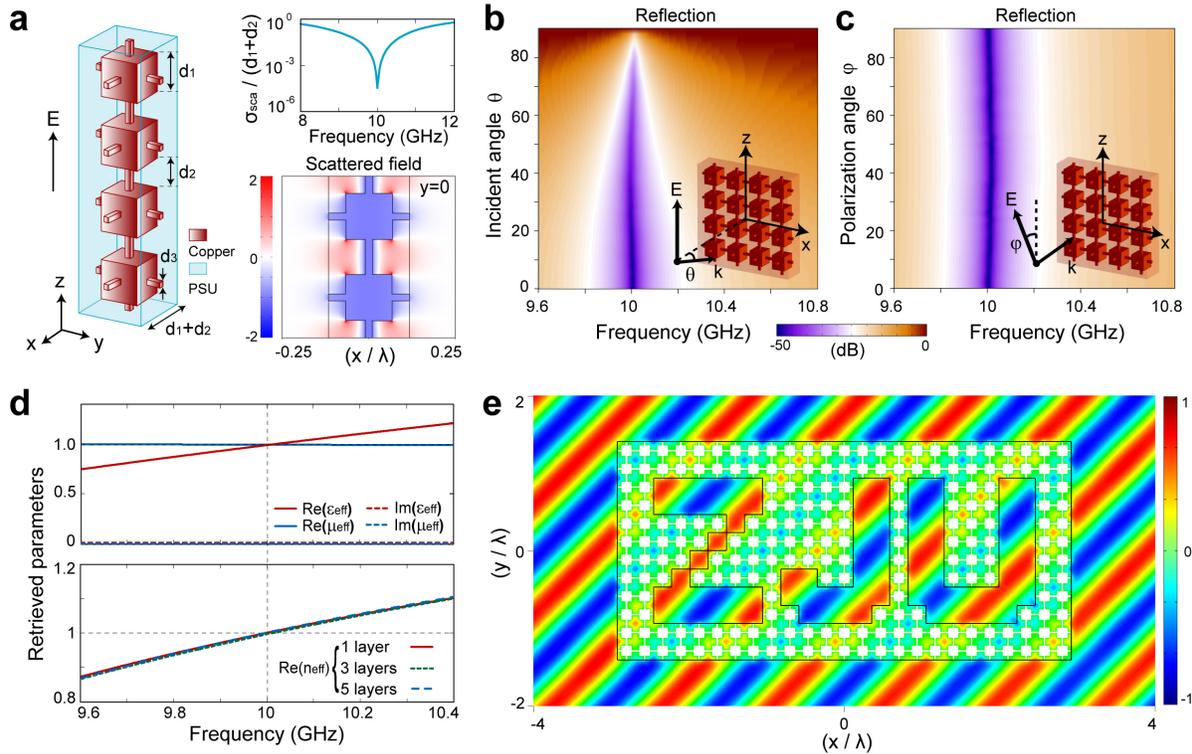

**Figure 2 | Numerical results of the invisible material made of corrugated metallic wires. a**, Left: the geometry of a modified wire, composed of corrugated conducting square wire embedded in a dielectric with $\varepsilon_r = 3$ and a loss tangent of 0.0013. Top right: the normalized scattering width ($5\times10^{-5}$ at 10 GHz) under a normal plane wave incidence along the *x*-axis with the unit-amplitude electric field polarized along the *z*-axis. Bottom right: the scattered electric field around the corrugated wire at 10 GHz. **b,** Simulated

reflectance spectra ($S_{11}$) for a one-layer of wires with respect to the angle of incidence. **c,** Simulated reflectance spectra with respect to the polarization angle. **d,** Retrieved effective parameters of the solid slab composed of closely arranged corrugated wires with different layer thickness. For the one-layer slab, $\varepsilon_{eff}$ = 0.9999+0.006i, $\mu_{eff}$ = 1 and the refractive index $n_{eff} \approx$ 1+0.003i at 10 GHz. **e,** Steady-state electric field distribution under an oblique plane wave incidence at 10 GHz upon a rectangular bulk composed of the designed wires with "ZJU"-shaped air cavity. The electric field is polarized along the z-axis.

For ease of fabrication, we modify the cylinders in the wire into cubes in Fig. 2a. We connect the cubes with thin square-shaped rods symmetrical in x, y and z directions. This makes the original wire structure isotropic in the 3D space, which removes the previous constrain that the field polarization has to be vertical. This conducting skeleton is embedded in a low-loss dielectric. Such a modified construction, still being sub-wavelength, has no qualitative change in its scattering properties from the corrugated wires studied analytically in Fig. 1. We performed full wave simulations on this rectangular wire structure using CST Microwave Studio™. The dimensions are $d_1$ = 4 mm, $d_2$ = 3 mm and $d_3$ = 0.6 mm. The conductor is copper with a conductance of $5.986 \times 10^7$ S/m, and the dielectric is polysulfone (PSU) with dielectric constant of 3 and loss tangent of 0.0013. Shown in Fig. 2a, the invisible frequency occurs at around 10 GHz with a normalized scattering width as low as $5 \times 10^{-5}$. The scattered electric field (difference between fields with and without the wire) is almost all localized inside the wire, consistent with the near zero scattering width. The opposite phase at different sections along the wire leads to the cancellation of the outgoing waves in the far field. We note that this structure has a low loss at the invisible frequency that is spectrally far away from the resonances.

When the wires are packed into a single 2D plane as the inset in Fig. 2b, the reflection spectra off the mesh sheet shows hardly any dispersion in either the polarization direction or incident angle, as long as the incident electric field is parallel to the sample plane (*S*-polarizated). The reflection is lower than -45 dB for normal incidence and remains below -30 dB for the incident angle of 80°. The performance is also independent of the polarization angle as shown in Fig. 2c, a result of its nearly isotropic in-plane geometry. Again, we show the effective constitutive parameters of this layer of wires in the top plot of Fig. 2d. At 10 GHz, $\varepsilon_{eff}$ = 0.9999+0.006i and $\mu_{eff}$ = 1. Accordingly, the real part of its effective refractive index is almost unity and it is nearly independent of the number of layers as shown in the bottom of Fig. 2d. So we can conclude that arbitrary arrangements of this mesh will be invisible as long as the electric field is parallel to all the interfaces within the beam width. To further illustrate this unique air-like material, we performed full wave simulations on a network of wires with selected wires removed to represent "ZJU" (Zhejiang University.) shown in Fig. 2e. Under an oblique incidence of plane wave at 10 GHz, the steady-state total electric fields in air stay undisturbed, showing a perfect invisibility. Animation of the electric field propagation can be found in **Supplementary Movies**.

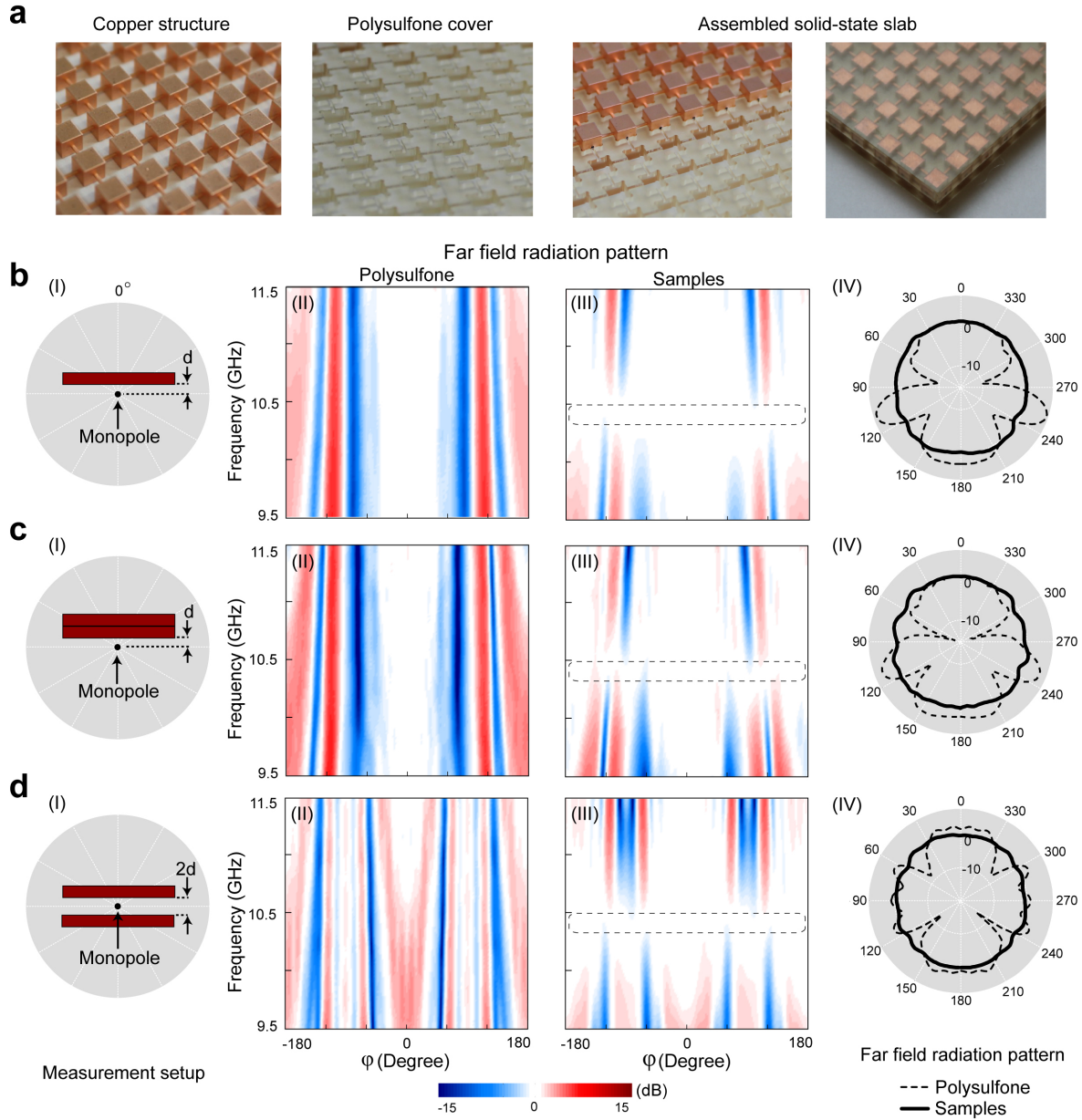

**Figure 3 | Experimental measurements of the fabricated samples. a**, Slab samples composed of closely arranged corrugated wires. **b,** Experimental far field radiation pattern measured for a single layer of slab. (I) shows the experimental setup, (II) and (III)

shows the measured field amplitude for an identical PSU slab and the sample slab, and (IV) shows the radiation pattern. **c,** Measurement for two stacked layers. **d,** Measurement for two spaced slabs. The radiation patterns in polar coordinates are plotted with data at 10.4 GHz.

As shown in Fig. 3a, the sample was fabricated by sandwiching the copper connected-cubes between two pieces of PSU covers. The dimensions of the samples are $217 \times 252 \times 7$ mm³ (31×36×1 in periods). To make the copper structure, we first 3D-print a plastic array of the connected cubes using stereolithography (material: Accura®60). Then a metal sputtering process was used to coat the surfaces of the plastic array with a 50 um of copper film that is well above the skin depth (0.64 um) at 10 GHz. The two PSU cover layers were machined with grooves and square openings so the copper structure can be embedded tightly inside.

In the measurements, three sets of sample configurations were studied. In the first configuration (Fig. 3b), one layer of the assembled slab was placed on a rotation stage spinning around a small monopole antenna with a separation distance of *d* from the slab, shown in panel (I). A wide band signal was fed into the monopole, and a wide band receiving lens antenna was placed at the other side to detect the far field radiation patterns by measuring transmission amplitudes $S_{21}$ (S-parameter) using an Agilent's E8361A network analyzer. For comparison, a reference measurement was performed by replacing the sample with a PSU slab of the same size. Shown in panel (II), the radiation pattern of the PSU sample is strongly directional for all frequencies. But for the designed sample shown in panel (III), there exists a frequency range around 10.4 GHz where the radiation pattern is almost isotropic within a 7.2% relative bandwidth where the scattering amplitudes are less than ±1 dB. For a direct comparison in panel (IV), we plot the transmission amplitudes in angular polar coordinates for both the sample and the reference at 10.4 GHz, validating the omnidirectional invisibility of the sample. Although fabrication imperfections inevitably degrade the performance and shift the operating frequency from 10 GHz to 10.4 GHz, the measurement results agrees with our analytic and numerical results.

In the second configuration (Fig. 3c), two slab samples were stacked together as a thicker one. Equivalent sets of measurements were performed as those for the first configuration. In the third configuration (Fig. 3d), the two slab samples were separated on the two sides of the source antenna. In both configurations, similar results were obtained as those of the first configuration. The above results confirm that the fabricated sample is omnidirectionally invisible regardless of its geometry.

In conclusion, we have demonstrated the ability to construct invisible microwave materials out of corrugated wires with record-low scattering width. Our analytical analyses, numerical simulations and experimental measurements are all consistent. We hope this work will inspire new technological applications, one important example being the construction of perfect antenna radomes. Objects can also be cloaked inside the metallic cubes. The proposed approach is simple, robust, and scalable to higher frequencies. Based on the general ability to control the frequency dispersions by multiple resonant structures[26], it should be possible to design wider-bandwidth materials invisible to both polarizations using our approach.

## Acknowledgements

We thank Zhiyu Wang, Hongshen Chen, Yichen Shen and Chia Wei Hsu for discussions. This work is supported by the NSFC under grants 61401393 and 61131002, and the China Postdoctoral Science Foundation under grant 2014M550325. J.D.J. was supported in part by the U.S. Army Research Office through the Institute for Soldier Nanotechnologies under contract W911NF-13-D-0001. L.L. was supported in part by the Materials Research Science and Engineering Center Program of the NSF under award DMR-1419807. M.S. and L.L. (analysis and reading of the manuscript) were supported in part by the MIT Solid-State Solar-Thermal Energy Conversion Center and Energy Frontier Research Center of DOE under grant DE-SC0001299.

## Corresponding author:

ranlx@zju.edu.cn and linglu@mit.edu